\documentstyle[12pt]{article}
\catcode`@=11
\def\@sim#1#2{\setbox0=\hbox{$\sim$}\lower.9\ht0\vbox{\baselineskip0pt
          \lineskip0.1ex\ialign{$\m@th#1\hfill##\hfill$\crcr#2\crcr
          \sim\crcr}}}
\def\lsim{\mathrel{\mathpalette\@sim<}}
\def\gsim{\mathrel{\mathpalette\@sim>}}
\catcode`@=12
\def\d{\displaystyle}
\newdimen\tdim

\catcode`\@=11

\def\@maketitle{\newpage   
 \null
 \vspace*{-1\headsep}      
 \vspace*{-1\headheight}
 \vspace*{-24pt}
 \begin{flushright}{\large       
   { \preprintno} \\ \@date}
 \end{flushright}
 \vskip \headsep     
 \vskip \headheight
 \bigskip
 \begin{center}            
{\LARGE \@title \par}
   \vskip 2em
   {\large
     \lineskip .5em
     \begin{tabular}[t]{c}\@author
     \end{tabular}\par}
   \vskip 1em
 \end{center}
 \par
 \vskip 1.5em}

\newcommand{\preprintno}{preprint number here}   

\def\abstract{\if@twocolumn
\section*{Abstract}
\else                         
\begin{center}
{\bf Abstract\vspace{-.5em}\vspace{0pt}}
\end{center}
\quotation
\fi}
\def\endabstract{\if@twocolumn\else\endquotation\fi}

\def\appendix{\par
    \setcounter{section}{0}
    \setcounter{subsection}{0}
    \renewcommand{\theequation}{\Alph{section}.\arabic{equation}}
    \setcounter{equation}{0}
}

\@addtoreset{equation}{section}
\def\theequation{\arabic{section}.\arabic{equation}}

\def\ps@columns{%
 \if@twocolumn
  \let\@mkboth\@gobbletwo
  \def\@oddhead{}\def\@evenhead{}
  \def\@oddfoot%
   {\rm\hfil\thepage\stepcounter{page}\hskip.5\textwidth\thepage\hfil}
  \let\@evenfoot\@oddfoot
 \else
 \ps@plain
 \fi
}

\catcode`\@=12

\makeatletter \input art10.sty \makeatother
\def\starttext{\twocolumn}
\def\tblvert{\vspace{-0.10in}}

\typein{*** or hit [return] for landscape mode(11x8.5) ***}

\newcommand{\beq}{\begin{equation}}
\newcommand{\eeq}{\end{equation}}

\newcommand{\remove}[1]{}

\renewcommand{\theequation}{\thesection.\arabic{equation}}
\begin{document}

\title{Effects of Top Compositeness \thanks{Research
supported in part by the National Science Foundation under Grant
\#PHY-9218167 and in part by Deutsche Forschungsgemeinschaft.}}

\author{
      Howard Georgi \thanks{georgi@huhepl.harvard.edu} \\
      Lev Kaplan \thanks {kaplan@huhepl.harvard.edu} \\
      David Morin \thanks {morin@huhepl.harvard.edu} \\
      Andreas Schenk \thanks {schenk@huhepl.harvard.edu} \\
      Lyman Laboratory of Physics \\
      Harvard University \\
      Cambridge, MA 02138}
\date{\today}

\renewcommand{\preprintno}{HUTP-94/A035\\ hep-ph/9410307}

\begin{titlepage}

\maketitle

\def\thepage {}        

\begin{abstract}
We investigate the effects of top quark compositeness on various
physical parameters, and obtain lower limits on the compositeness
scale from electroweak precision data.  We consider corrections
to top quark decay rates and other physical processes. Our results
depend sensitively on whether the left-handed top is
composite. A considerable enhancement of $t \bar t$ production is
possible if only the right-handed top is composite.
\end{abstract}

\end{titlepage}

\starttext 
\pagestyle{columns} 
\pagenumbering {arabic} 

\section{Introduction}

The CDF collaboration at FNAL has recently found direct evidence for a top
quark with a mass of about 175~GeV~\cite{CDF}. Due to its large mass,
the top couples more strongly than the other quarks to the longitudinal
electroweak gauge bosons. This makes it potentially interesting as a probe of
physics beyond the standard model of electroweak interactions. Within the
standard model, the $\rho$-parameter and the decay width $\Gamma(Z
\rightarrow b\bar b)$, for example, receive corrections from top loops which
depend quadratically on the mass of the top quark. Other low energy parameters
depend logarithmically on this mass. Such quantities are likely to be affected
by the introduction of a top quark with non-standard properties. Because of
its large mass, the top quark also seems the most likely candidate of all the
known particles for interacting with physics beyond the standard model. The
preliminary evidence in Ref.~\cite{CDF} for an enhanced top production rate is
a tantalizing indication of how such physics might show up in more definitive
experiments. Finally, if one assumes that the scale $\Lambda$ characterizing
the new physics is large, then its effects, which are suppressed by powers of
$1/\Lambda$, are most likely to be seen in processes involving the top.

In this work, we will investigate the effects of a composite top quark on
physics accessible at current particle accelerators.
In particular, we will consider two scenarios: one where only the
right-handed top is composite, and the other where the left-handed
doublet containing the top is composite (in this case it will not make any
qualitative difference whether the right-handed top is composite as well). Our
primary concern is the question of whether the success of the standard model
in predicting the properties of the $Z$ and $W$ already so strongly
constrains the physics of the $t$ that any new physics would not be
observable at presently accessible energies. Somewhat to our surprise, we
find that the answer is no.

The possibility of non-standard couplings of the top to the electroweak gauge
bosons has recently been analyzed in Ref.~\cite{Yuan}.\footnote{For earlier
works on non-standard $t$ couplings, see \cite{earlier}.} The authors of
Ref.~\cite{Yuan} assume that the non-standard properties of the top quark are
associated with the electroweak symmetry breaking sector, with new physics at
a scale of a few TeV. They obtain limits on the dimensionless coefficients of
non-standard dimension-four operators rather than bounds on the energy scale
of the new physics. The scheme that we discuss is potentially more
interesting.  We will assume that the physics of top compositeness does not
directly involve the electroweak symmetry breaking mechanism. We can then
consider the possibility that the top could participate in new physics {\it at
a scale significantly below 1~TeV.} We will show that if only the right-handed
top participates in the new physics, this is consistent with the bounds on the
new physics from precise electroweak tests at LEP and SLC. However, if the
scale of the new physics is this low, other properties of the top besides the
coupling to electroweak gauge bosons will be affected, possibly leading to
measurable deviations from standard model predictions for production rates,
decay widths, and so on.

To think seriously about new physics at scales below 1~TeV, we will have to
make some assumptions about the flavor structure of the new physics in order
to satisfy experimental bounds on flavor changing neutral current processes
and CP violation. We will do this without apology. These are issues that any
compositeness theory must address. In this note, we concentrate on the
electroweak physics that must be present in any such theory.

We will use the technique of effective field theory, applying naive
dimensional analysis to estimate the coefficients of various terms in the
effective theory Lagrangian. We obtain a lower bound on the compositeness
scale $\Lambda$ using the electroweak precision parameters measured at LEP:
the oblique corrections $S,\ T,\ U,$ and the partial decay width $\Gamma(Z
\rightarrow b \bar b)$. We will also analyze the effects of compositeness on
various physical processes, including $t \bar t$ production and top decays.

Because our analysis involves only the low energy theory, valid below the
compositeness scale, we do not need to discuss the details of the new strong
interactions in which the $t$ participates. This is just as well, because we
do not have any completely satisfactory proposals for this physics. All
schemes that we know of have unattractive features such as fundamental
spinless fields at some scale. Our view is that if compositeness is realized
in nature, it will involve properties of strongly interacting chiral gauge
theories that we do not now understand or even imagine. Exciting new physics
of this kind might show up first in the properties of the $t$ that we discuss
in this note.

\section{Theory}

We assume that the effects of a composite top can be described by an effective
theory~\cite{efftheory} below the compositeness scale $\Lambda$. The leading
part of the effective theory Lagrangian is the standard model of electroweak
interactions. We will further assume that the scale $\Lambda$ is larger than
the mass of the standard model Higgs; therefore the Higgs field is present in
the effective theory Lagrangian. For a very heavy Higgs this Lagrangian would
contain additional terms which arise from ``integrating out" the Higgs
particle. In the heavy-Higgs scenario, our basic conclusions would be
unchanged.

Corrections coming from the underlying theory are described by
operators of dimension six and higher which are suppressed by powers of
$1/\Lambda$~\cite{Wyler}, where $\Lambda$ is defined to be the scale at
which
the low energy theory breaks down. In the following analysis, we will only
consider the effects of dimension six operators, which are suppressed by
$1/\Lambda^2$. The effects of higher dimension operators, which are suppressed
by higher powers of $1/\Lambda$, are expected to be subleading, and are
neglected in our analysis.

We now need a rule to estimate the sizes of the coefficients
of nonrenormalizable operators in the effective theory. The rule we
will use is that of naive dimensional analysis (NDA)~\cite{NDA}.
In analogy to QCD, we introduce a second scale, $f$, such that $1/f$
characterizes the amplitude for creation of a field that interacts
strongly with the underlying physics. The convergence of the loop
expansion in the effective theory for energies below $\Lambda$ requires
\begin{equation} \label{fbound}
{\Lambda\over 4\pi f}\equiv \rho \lsim 1\ .
\end{equation}
In our case, the only field in the low energy theory with these new strong
interactions
is the composite top. The standard model gauge fields interact weakly
with the physics of top compositeness (in the scenario where only the
right-handed top is composite, the ${\rm SU}(2)$ gauge fields need not
interact with the constituents of the top at all). The remaining standard
model fields, namely the Higgs doublet and the non-composite fermions,
do not interact at all with the top constituents.

The rules for estimating the coefficients of terms involving the composite
top and covariant derivatives are as follows. Each power of the composite
top field is accompanied by a factor of $1/f$. Furthermore, by analogy
with QCD, there is an overall normalization factor of $f^2$. Finally, one
has to include an appropriate power of $1/\Lambda$ to obtain the correct
dimension. We assume that with this normalization, all dimensionless
coefficients are of order one.
At dimension six, there are three types of terms involving fields
which interact with the underlying physics. The terms involving
four composite top fields are suppressed by $1/f^2$, while those involving
two composite fields and three covariant derivatives are suppressed by
$1/\Lambda^2$. Finally, terms with six powers of covariant derivatives
are suppressed by $f^2/\Lambda^4$.

Because the scalar doublet and the non-composite fermions do not interact
directly with the underlying physics, effective operators involving these
fields are generated only as counterterms to loops with insertions of
the operators described above. The finite parts of the counterterm operators
are of the same order as the finite parts of the loop
diagrams with which they are associated.

Note that the definition of the effective theory Lagrangian is ambiguous.
For example, one can use equations of motion to change the relative sizes of
the coefficients of various terms.
Thus, more precisely, our assumption is that this Lagrangian can be written in
such a way that the rules of NDA are satisfied.

We will also assume below that the new physics is CP-conserving. If we do not
make this assumption, the new physics will give rise to CP-violating 3-gluon
couplings in the low energy theory which will contribute to the neutron
electric dipole moment~\cite{cpweinberg}. CP violation of ${\cal O}(1)$ in
the physics of top compositeness would put such a strong
constraint on the scale of the new physics that there would be no possibility
of seeing interesting effects at accessible scales. Thus in the effective low
energy theory, we do not include CP violating couplings except for those that
arise in the standard model from the physics of quark mass generation and the
CKM matrix.

For the right-handed case, then, the leading corrections to all physical
parameters which we consider
are determined by the following operators in the effective Lagrangian,
with coefficients of order one:
\def\Dslash{\not{\!\!D}}
\begin{equation} \label{Refffour}
{1 \over f^2} \overline{T_R} \gamma^\mu T_R \overline{T_R} \gamma_\mu T_R \
\end{equation}
and
\begin{displaymath}
{i \over\Lambda^2} \overline{T_R} D^2 \Dslash T_R \ , \
{i \over\Lambda^2} \overline{T_R} \Dslash D^2 T_R \ , \
{i \over\Lambda^2} \overline{T_R} D_\mu \Dslash D^\mu T_R \ ,
\end{displaymath}
\begin{equation} \label{Reffdef}
g' {i \over\Lambda^2} \overline{T_R} B_{\mu\nu}  D^\mu \gamma^\nu T_R \ , \
g_s {i \over\Lambda^2} \overline{T_R} G_{\mu\nu}  D^\mu \gamma^\nu T_R \ ,
\end{equation}
where
\begin{equation}  \label{admixed1}
T_R = t_R + \theta^c_R c_R + \theta^u_R u_R \ .
\end{equation}
In this model we admit small admixtures of the charm and up
quarks to the composite particle. Out of the five operators in
Eq.~(\ref{Reffdef}), only four are linearly independent.
Note that we have omitted
all dimension six operators which arise as counterterms and
produce corrections of the same order as loop diagrams with
insertions of terms in Eqs.~(\ref{Refffour},\ref{Reffdef}). Since we are only
interested in order of magnitude estimates, we can neglect the
contributions of these operators.

For the left-handed case, the corresponding terms are given by
\begin{equation} \label{Lefffour}
{1 \over f^2} \overline{\Psi} \gamma^\mu \Psi \overline{\Psi} \gamma_\mu \Psi
\ ,
\end{equation}
and
\begin{eqnarray} \label{Leffdef}
& \d {i \over\Lambda^2} \overline{\Psi} D^2 \Dslash \Psi \ , \
\d {i \over\Lambda^2} \overline{\Psi} \Dslash D^2 \Psi \ ,   \nonumber \\
& \d {i \over\Lambda^2} \overline{\Psi} D_\mu \Dslash D^\mu \Psi \ , \
\d g' {i \over\Lambda^2} \overline{\Psi}
B_{\mu\nu} D^\mu \gamma^\nu \Psi \ , \\
& \d g {i \over\Lambda^2} \overline{\Psi}
W_{\mu\nu} D^\mu \gamma^\nu \Psi \ , \
\d g_s {i \over\Lambda^2} \overline{\Psi}
G_{\mu\nu} D^\mu \gamma^\nu \Psi \ ,
\nonumber
\end{eqnarray}
and
\begin{equation} \label{Leffgauge}
g^2{f^2\over\Lambda^4} D_\lambda W_{\mu\nu}^a D^\lambda W^{a\mu\nu} \ , \
g^3{f^2\over\Lambda^4} \epsilon_{abc} W_{\lambda}^{a\mu} W_\mu^{b\nu}
W_\nu^{c\lambda} \ , \
g^2g'{f^2\over\Lambda^4} W_{\lambda}^{a\mu} W_\mu^{a\nu} B_\nu^{\lambda} \ ,
\end{equation}
where
\begin{equation}
\Psi=\left(\begin{array}{c}{ T_L}\\ {B_L} \end{array} \right)
\end{equation}
and $g$ and $g'$ are the SU(2) and U(1) gauge couplings, respectively.
Again, one of the operators in Eq.~(\ref{Leffdef}) may be omitted.
In terms of the mass eigenstates, we have
\begin{eqnarray}  \label{mixmass}
& & T_L = t_L + \theta^c_L c_L + \theta^u_L u_L  \ , \\
& & B_L = b_L +  \theta^s_L s_L +
\theta^d_L d_L \nonumber \ .
\end{eqnarray}
The condition that $\Psi$ be a linear combination of electroweak doublets,
which is necessary because the new physics preserves electroweak SU(2),
is~\cite{Conrad}
\begin{equation}
\theta^s_L=\theta^c_L + V_{ts}\,,\quad\quad
\theta^d_L=\theta^u_L + V_{td}\,,
\label{kmcondition}
\end{equation}
where $V_{ts}$ and $V_{td}$ are the CKM matrix elements, and terms of higher
order in the mixing angles are neglected.
In Eq.~(\ref{Leffgauge}), we have omitted terms involving only the
${\rm U}(1)$ and ${\rm SU}(3)$ gauge fields, because they will only contribute
to physical parameters of interest at higher orders.

The effects of top compositeness on various physical parameters are summarized
in Table~1.

\def\mtl{{m_t^2\over\Lambda^2}}
\def\mzl{{M_Z^2\over\Lambda^2}}
\def\mtf{{m_t^2\over(4\pi f)^2}}
\def\order#1{ {\cal O} \Big( #1 \Big) }

\begin{table}[htbp]
\begin{center}
\begin{tabular}{|c|c|c|} \hline
Parameter & Composite left-handed top & Composite right-handed top
\\ \hline
\tblvert &&\\
$S$ & $\d \mtl S_{SM}$ & $\d{\mtl} S_{SM}$ \\
\tblvert &&\\
$T$ & $\d{\mtl} T_{SM}$ & $\d{\mtl} T_{SM}$ \\
\tblvert &&\\
$U$ & $\d{\mtl} U_{SM}$ & $\d{\mtl} U_{SM}$ \\
\tblvert &&\\
$\Gamma(Z \rightarrow c \bar c)$ & $\d |\theta_L^c|^2 \mtf
 \Gamma^{\rm tree}_{SM}(Z \rightarrow c \bar c) $&
$\d |\theta_R^c|^2 \mtf
 \Gamma^{\rm tree}_{SM}(Z \rightarrow c \bar c)$ \\
\tblvert &&\\
$\Gamma(Z \rightarrow b \bar b)$ & $ \d \mtf
 \Gamma^{\rm tree}_{SM}(Z \rightarrow b \bar b) $&
$\d{\mtl} \Gamma^{\rm loop}_{SM}(Z \rightarrow b \bar b) $\\
\tblvert &&\\
$\Gamma(Z \rightarrow s \bar s)$ & $\d |\theta_L^s|^2 \mtf
 \Gamma^{\rm tree}_{SM}(Z \rightarrow s \bar s) $&
$ \d  |\theta_R^c|^2 {\mtl}
 \Gamma^{\rm loop}_{SM}(Z \rightarrow s \bar s) $\\
\tblvert &&\\
$\Gamma(t \rightarrow Wb)$ &
$\d \mtl
\Gamma^{\rm tree}_{SM}(t \rightarrow Wb) $ & $ \d {\mtl}
 \Gamma^{\rm loop}_{SM}(t \rightarrow Wb)$ \\
\tblvert &&\\
$\Gamma(t \rightarrow Zc)$ &
$ \d {|\theta^c_L g_L|^2\over 2\pi}
        {m_t^7\over  M_Z^2 \Lambda^4} $ &
$ \d {|\theta^c_R g_R|^2\over 2\pi}
        {m_t^7\over  M_Z^2 \Lambda^4} $  \\
\tblvert &&\\
$\Gamma(b \rightarrow \gamma s)$ &
$ \d {|\theta^s_L| \over |V_{ts}|}
 {16 \pi^2 \over 2g^2} {M_W^2 \over \Lambda^2}
 \Gamma_{SM}(b \rightarrow \gamma s) $ &
$ \d {m_t^2 \over \Lambda^2} \Gamma_{SM}(b \rightarrow \gamma s) $ \\
\tblvert &&\\
$m_{K_L}\! - m_{K_S}$  & $\d|\theta^s_L \theta^d_L |^2
{{f_K^2\over f^2}} m_K$ & --- \\
\tblvert &&\\
$m_{D_1^0}\! - m_{D_2^0} $ & $\d|\theta^c_L \theta^u_L |^2
{{f_D^2\over f^2}} m_D$
& $\d|\theta^c_R \theta^u_R |^2
{{f_D^2\over f^2}} m_D$  \\
\tblvert &&\\
\raisebox{0in}[0in][0.2in]{$m_{B_1^0}\! - m_{B_2^0} $}
& $\d|\theta^d_L|^2
{{f_B^2\over f^2}} m_B  $& --- \\
\hline
\end{tabular}
\caption{Effects of top compositeness on various physical parameters.
All entries are estimates, with unknown numerical coefficients of order one.
Only the leading contributions are included in the table. $Z$-boson decays
with corrections of order $m_t^2/(4\pi f)^2$ also receive contributions
of order $M_Z^2/\Lambda^2$.
The latter may be leading if $4 \pi f /\Lambda \protect\gsim 2$.
Where possible, corrections are compared to tree or loop standard model
contributions.
$S_{SM}, T_{SM},$ and $U_{SM}$ are contributions to the oblique parameters
coming from top loops in the standard model.
The $\theta$ angles are defined
in Eqs.~(\protect\ref{admixed1}, \protect\ref{mixmass},
\protect\ref{kmcondition}),
and the couplings $g_{L,R}$ are defined in Eq.~(\protect\ref{gdef}).}
\end{center}
\end{table}

\section{Discussion and Conclusions}

We will proceed to discuss in turn each of the entries in the table.
The electroweak precision parameters $S,\ T,$ and $U$
receive contributions only at the one-loop level.
Note that in our notation, the
quantities $S_{SM},\ T_{SM},$ and $U_{SM}$ describe contributions to the
oblique corrections coming from top loops in the standard model. Since
our results are only
correct up to factors of order one, we do not differentiate between constant
terms and terms involving the logarithm of $(M_W/m_t)^2$. In this
approximation, our results can be cast into the form given in the table.

The S-parameter,
for example, will receive contributions from the diagrams in
Fig.~(1). Diagrams (1a) and (1b) represent top loops with insertions
of operators from Eqs.~(\ref{Reffdef},\ref{Leffdef}). In the
case of a left-handed composite top, there is also a diagram with
an effective operator insertion on the other vertex. These diagrams give
contributions of order $m_t^2/\Lambda^2$ compared to the standard model
effects, as given in the table. It is easy to see that this must be the
size of the leading correction, because by dimensional analysis a quantity
of dimension mass squared must enter in the numerator to counter the factor
of $1/\Lambda^2$ from the insertion, and $m_t$ is the largest mass scale
available. Diagram (1c) represents a top loop
with an insertion of the four-top vertex from Eqs.~(\ref{Refffour},
\ref{Lefffour}). This effect is suppressed by a factor of
$m_t^2/(4\pi f)^2$ instead. If the bound~(\ref{fbound}) is saturated,
then this contribution is of the same order as the ones from
diagrams (1a) and (1b), otherwise
it is smaller. Finally, diagram (1d) represents contributions from
the counterterms. The contributions to $T$ and $U$ come from diagrams
analogous to those in Figs.~(1b-d).

In the case of left-handed top compositeness,
the parameters $S,\ T$, and $U$ also receive contributions from loops with
insertions of operators from Eq.~(\ref{Leffgauge}). However, these
effects are at most of order $f^2 M_Z^2/\Lambda^4$ relative to the standard
model, and are negligible compared to the contributions discussed previously.
Effective operators involving only the gauge fields do not contribute to the
oblique corrections at tree level, because such operators are
${\rm SU}(2)$-symmetric.

In the case of left-handed top compositeness, the tree contribution
to $Z$ decays is given by diagram (2a). It contains an effective vertex
coming from Eq.~(\ref{Leffdef}). This effect, of order
$M_Z^2/\Lambda^2$ relative to the standard model contribution, may be
subleading
as compared to the loop diagram shown in Fig.~(2b). The latter has an
effective four-quark insertion from Eq.~(\ref{Lefffour}), and is suppressed
relative to the standard model by factor of $m_t^2/(4\pi f)^2$. In the
case of right-handed top compositeness, the contribution to
$\Gamma(Z \rightarrow c \bar c)$ is again given by the diagrams in
Fig.~(2). The other two $Z$ decays listed in the table receive leading
contributions from triangle diagrams with two internal top lines and
an insertion of an effective operator from Eq.~(\ref{Reffdef}).
If cancellations between the various diagrams occur, the actual
corrections might turn out to be even more suppressed. We will
see below that our estimate of the scale $\Lambda$ is not changed if such
cancellation indeed occurs.

The corrections to the decay rates of the top quark arise from
diagrams analogous to those in Fig.~(2). In this case, the contributions
from diagram~(2a) are suppressed by $m_t^2/\Lambda^2$, and are of the same
order as those from diagram~(2b) for $\Lambda \approx 4\pi f$.
Since the standard model contribution for the GIM-violating
decay $t \rightarrow Z c$ is extremely small, the table explicitly shows the
leading result for
this decay mode. The coupling constants $g_R$ and $g_L$ are given by
\begin{equation} \label{gdef}
g_R = -e{2\over 3} {s_\theta\over c_\theta} \ ,  \ \
g_L = {e \over 2 s_\theta c_\theta} (1 - {4 \over 3} s_\theta^2) \ ,
\end{equation}
where $s_\theta$ and $c_\theta$ are the sine and cosine of the Weinberg
angle, respectively.

The corrections to $\Gamma(b \rightarrow \gamma s)$ (this decay is
considered in Ref.~\cite{fujikawa} in the context of an alternative
extension to the standard model) arise in our case from Figs.~(3a)
and (3b), in the scenarios of left- and right-handed top compositeness,
respectively. Note that in the case of left-handed compositeness, there
is a tree-level contribution to this process, whereas the leading standard
model contribution is at the one-loop level. In the case of right-handed
top compositeness there are also diagrams similar to Fig.~(3b), but with
the operator insertion at or to the right of the photon vertex. Also notice
that in contrast to the analysis in Ref.~\cite{fujikawa}, we do not
include a composite right-handed bottom quark in either scenario, and
consequently we do not get an $m_t/m_b$ enhancement for this process.

The leading contribution to the mass splittings of the neutral mesons is given
by the formula
\begin{eqnarray}
\Delta m_P & = &{1\over 2 m_P} \langle P^0 | {\cal H} | \overline{P^0} \rangle
 \\        & = &{1\over 2 m_P} A_P \langle P^0 |
        {\cal O}^{\overline{q_1} q_2} | \overline{P^0} \rangle \ ,
\end{eqnarray}
where ${\cal O}^{\overline{q_1} q_2}$ is the appropriate four-quark operator
whose
matrix element is given by $f_P^2 m_P^2 $ up to some numerical factor of order
one. This matrix element contains the effects of strong interaction physics.
The quantity $A_P$ is the corresponding four-quark scattering amplitude. In
the
standard model it describes electroweak effects on the mass mixing. In our
case, this amplitude includes the effects of top compositeness as well.  The
leading contribution is determined by the four-top-quark operators, which are
suppressed by a factor of $1/f^2$.
These contributions to the mass splittings are further suppressed by
the small mixing angles.

Let us now proceed to obtain estimates on the scales $\Lambda$ and $f$.
We should stress that all these estimates are valid only up to factors
of order one. The constraints come mainly from the electroweak precision
parameters $S,\ T,\ U$, and $\gamma_b$ (the last one associated with
$Z \rightarrow b \bar b$ decay), a recent discussion of which can be
found in Ref.~\cite{Barbieri}. The constraints coming from the oblique
parameters $S,\ T,$ and $U$ do not discriminate between the left- and
right-handed cases. The combined estimate from these three parameters is
\begin{equation} \label{lrbound}
\Lambda_{L,R} > (1  \sim 2) m_t \ .
\end{equation}
The constraint from the parameter $\gamma_b$ is much stronger in the case
of left-handed top compositeness, requiring
\begin{equation} \label{lambound}
\Lambda_L \gsim \max(10 \rho \,m_t,10 M_Z) \ ,
\end{equation}
where $\rho$ is defined in Eq.~(\ref{fbound}).
The limit $\Lambda_L\gsim10M_Z$ comes from the contribution of the diagram in
Fig.~(2a), as discussed above. This will be the less stringent constraint
unless the bound (\ref{fbound}) is at least a factor of two away from being
saturated. In the right-handed case, on the other hand, the bound from
$\gamma_b$ is of the same order as that coming from the oblique corrections.
Also, in the right-handed scenario, the bound on $\Lambda_R$ coming from
$\Gamma(b \rightarrow \gamma s)$ is again of this order. In the case of
left-handed compositeness, constraints coming from
$\Gamma(b \rightarrow \gamma s)$ depend on the mixing angle $\theta^s_L$,
and will be discussed below.

Now we discuss limits on the mixing angles $\theta^u$ and $\theta^c$, which
are defined in Eqs.~(\ref{admixed1}, \protect\ref{mixmass},
\protect\ref{kmcondition}). The limits coming from the $Z$ decays are not very
stringent; therefore we will use constraints coming from neutral meson
mixings. A general remark is that if we are willing to fine tune the $\theta$
parameters, present bounds on neutral meson mixings yield no constraints. For
the case of right-handed $t$ compositeness, this is trivial, because
$\theta^c_R$ and $\theta^u_R$ can be tuned to zero, making the new physics
flavor conserving. However, this is not possible for the left-handed case,
because of (\ref{kmcondition}). Nevertheless, in this case, by tuning
$\theta^c_L$ and $\theta^d_L$ to zero, we can simultaneously eliminate the
compositeness contributions to  $D^0-\overline{D^0}$, $K^0-\overline{K^0}$,
and $B^0-\overline{B^0}$ mixings, and thus present experimental limits give no
constraints~\cite{Conrad}. However, this scenario has interesting implications
for $B_s^0-\overline{B_s^0}$ mixing, which will be dramatically different from
its standard model value if the scale of the new physics is small.

Below, we discuss the fine-tuning required more quantitatively. In this
discussion, we will take the
decay constants $f_D$ and $f_B$ to be roughly equal to $f_K$; this should
suffice because we are only interested in obtaining estimates up to factors of
order one.

In the case of right-handed top compositeness we obtain from
$D^0-\overline{D^0}$
mixing the limit $|\theta^u_R \theta^c_R| \lsim 2.5 \times 10^{-5}$,
using a value of $(200/4\pi)$~GeV for the
scale $f$ (we take $200$~GeV to be the lower limit implied by
Eq.~(\ref{lrbound})). If the bound~(\ref{fbound}) is not saturated,
the constraint
on the angles is weaker, for the same value $\Lambda \approx 200$~GeV.

In the left-handed case, we obtain from
$B^0-\overline{B^0}$ mixing
the limit $|\theta^d_L| \lsim 3 \times 10^{-4}$, using the value
$150$~GeV for $f$, the smallest value consistent with Eq.~(\ref{lambound}).
The data from $K^0-\overline{K^0}$ mixing yields
$|\theta^d_L \theta^s_L| \lsim 4 \times 10^{-5}$. Thus, we get
the bound $|\theta^s_L| \lsim 1 \times 10^{-1}$. Since the CKM
matrix $V_{td} \gsim 3 \times 10^{-3}$, some fine-tuning for the angle
$\theta^u_L$ is required. On the other hand, since $V_{ts} \gsim
3 \times 10^{-2}$, no fine-tuning is necessary for $\theta^c_L$.
The bound from $D^0-\overline{D^0}$ mixing,
$|\theta^u_L \theta^c_L| \lsim 1.5 \times 10^{-3}$, does not provide any
additional information.

One may avoid fine-tuning for $\theta^u_L$ by choosing the following
much stronger bound, $4 \pi f_L \gsim 100 m_t$. (However, the bound on
$\Lambda_L$ as given in Eq.~(\ref{lambound}) need not be changed
by as large a factor as long as we allow the bound in
Eq.~(\ref{fbound}) not to be
saturated.) Using this stronger bound on $f$, we obtain
$|\theta^u_L| \lsim 3 \times 10^{-3}$, while the bound on $\theta^c_L$
is unchanged.

Using the above value for $\theta^s_L$, the decay rate
$\Gamma(b \rightarrow \gamma s)$ would yield a slightly stronger constraint
on $\Lambda_L$, compared to Eq.~(\ref{lambound}). However, lowering
the bound on $\theta^s_L$ by a factor of ten, to $|\theta^s_L| \lsim
1 \times 10^{-2}$, would allow us to retain the bound (\ref{lambound})
on the left-handed compositeness scale. In this case, we would no
longer be able to make $\theta^c_L$ arbitrarily small because of the
constraint given by Eq.~(\ref{kmcondition}). However, this would still
be consistent with presently available data on flavor-changing neutral
currents.

Now that we have an estimate on the scale $\Lambda$, we can use its value to
discuss the consequences of top compositeness on the decay widths $\Gamma(t
\rightarrow W b)$ and $\Gamma(t \rightarrow Zc)$, as well as on $t\bar t$
production. In the left-handed case, the correction to the standard model
prediction for $\Gamma(t \rightarrow W b)$ depends on the bound we choose
for $\Lambda$. In the cancellation scenario, where $\Lambda \gsim 1$~TeV
(the smallest value consistent with Eq.~(\ref{lambound})),
a correction of about 5\% is possible. In the other scenario, with
$4 \pi f \gsim 100 m_t$, the
correction is much smaller, by an amount which depends on the ratio we
choose for $4 \pi f/\Lambda$. In the
right-handed case, the constraint on $\Lambda$ is much weaker; nevertheless,
the
corrections to the standard model are only of the order of 0.1\% because the
leading
contribution appears only at the one loop level. The decay width $\Gamma(t
\rightarrow Zc)$ is suppressed relative to $\Gamma(t \rightarrow W b)$ by a
factor of $|\theta^c|^2 (m_t^2/\Lambda^2)^2$.
In the right-handed case, there is an
additional suppression by a factor of $\sin^2\theta_W$.

Recent findings on the top quark search~\cite{CDF}
suggest that there might be an
enhancement of $t \bar t$ production as compared to the standard model
prediction. In the effective theory the leading correction to the standard
model production amplitude grows like $p^2/\Lambda^2$ above threshold.
Using our estimate for the scale $\Lambda$, we see that in
the case of right-handed top compositeness there can indeed be sizable
corrections to the cross section for this process if $\Lambda$ is about
$2 m_t$. However, in this case, the leading correction is not dominant, and
the effective theory cannot be used to
predict the quantitative behavior of the amplitude above threshold.

In the left-handed case, the
contribution is considerably smaller due to the more stringent lower bound on
$\Lambda$.

In addition to the effects on the electroweak parameters, top compositeness
will also affect the physics of strong interactions, such as jet production
rates. In our model, these effects will contribute only indirectly through
quark loops. Thus the lower bound on the top compositeness scale that we
obtain from QCD is smaller by a factor of $16\pi^2$ than the bound on the
gluon compositeness scale, which was found to be about 7000~GeV (using
our normalization conventions) in
Ref.~\cite{gluon}. Therefore strong interaction physics provides a weaker
bound on the top compositeness scale than electroweak physics.

In our estimates on $\Lambda$ we have neglected the logarithmic running of
the effective coupling constants. In the interesting scenario of
small $\Lambda$, which is possible only in the case of right-handed top
compositeness, the running associated with the change of scale from
$\Lambda$ to $m_t$ is negligible.

To summarize, we find that in the case of left-handed top compositeness, the
effects on experiments at currently accessible energies are not very
interesting, due to the high scale $\Lambda$. In the right-handed case,
however, the compositeness scale can be much lower, allowing for sizable
corrections to standard model predictions. In particular, $t \bar t$
production can be greatly enhanced in this scenario.

\section*{Acknowledgements} HG is grateful to Lisa Randall for discussions of
compositeness. Research
supported in part by the National Science Foundation under Grant
\#PHY-9218167 and in part by Deutsche Forschungsgemeinschaft.

%
%
\newcommand{\artref}[4]{{#1} {\it #2} {#3} #4}
\newcommand{\cartref}[5]{{\sc #1}, {\it #2},  #3 {\bf #4}, #5}
\newcommand{\bookref}[2]{{\sc #1}, #2}

\section*{Figure Captions}

\begin{enumerate}
\item Feynman diagrams contributing to the $S$, $T$ and $U$ parameters.
\item Feynman diagrams contributing to the processes $Z\rightarrow
q\overline{q}$.
\item Feynman diagrams contributing to the process $b \rightarrow \gamma s$.
\end{enumerate}


\section*{Figures}

\newdimen\tdim
\tdim=1.5pt
\input prepictex
\input pictex
\input postpictex
\newcommand{\stpltsmbl}{\setplotsymbol ({\small .})}
\newcommand{\tarrow}{\arrow <5\tdim> [.3,.6]}
\newcommand{\EX}{\beginpicture
\setcoordinatesystem units <\tdim,\tdim>
\setplotsymbol ({\Large .})
\plot 5 5 -5 -5 /
\plot 5 -5 -5 5 /
\linethickness=0pt
\putrule from -5 0 to 5 0
\putrule from 0 -5 to 0 5
\endpicture}
\newbox\phrd
\setbox\phrd=\hbox{\beginpicture
\setcoordinatesystem units <\tdim,\tdim>
\stpltsmbl
\setquadratic
\plot
0 0
2.5 -3
5 0
7.5 3
10 0
/
\endpicture}
\def\photonrd #1 #2 *#3 /{\multiput {\copy\phrd}  at
#1 #2 *#3 10 0 /}

$$\beginpicture
\setcoordinatesystem units <\tdim,\tdim>
\stpltsmbl
\circulararc 360 degrees from 20 0 center at 0 0
\photonrd -50 0 *2 /
\photonrd 20 0 *2 /
\tarrow from 2 20 to -2 20
\tarrow from -2 -20 to 2 -20
\put {\large $t$} at 0 30
\put {\large $t$} at 0 -30
\put {\large $W_3$} [r] at -55 0
\put {\large $B$} [l] at 55 0
\put {\EX} at 20 0
\linethickness=0pt
\put {\large Fig. 1a} at 0 -50
\putrule from 0 -45 to 0 45
\endpicture$$

$$\beginpicture
\setcoordinatesystem units <\tdim,\tdim>
\stpltsmbl
\circulararc 360 degrees from 20 0 center at 0 0
\photonrd -50 0 *2 /
\photonrd 20 0 *2 /
\tarrow from -2 -20 to 2 -20
\put {\large $t$} at 0 30
\put {\large $t$} at 0 -30
\put {\large $W_3$} [r] at -55 0
\put {\large $B$} [l] at 55 0
\put {\EX} at 0 20
\linethickness=0pt
\putrule from 0 -45 to 0 45
\put {\large Fig. 1b} at 0 -50
\endpicture$$

$$\beginpicture
\setcoordinatesystem units <\tdim,\tdim>
\stpltsmbl
\circulararc 360 degrees from 20 0 center at 0 0
\circulararc 360 degrees from 20 40 center at 0 40
\photonrd -50 0 *2 /
\photonrd 20 0 *2 /
\tarrow from 2 60 to -2 60
\tarrow from -2 -20 to 2 -20
\put {\large $t$} at 0 70
\put {\large $t$} at 0 -30
\put {\large $W_3$} [r] at -55 0
\put {\large $B$} [l] at 55 0
\put {\EX} at 0 20
\linethickness=0pt
\putrule from 0 -45 to 0 85
\put {\large Fig. 1c} at 0 -50
\endpicture$$

$$\beginpicture
\setcoordinatesystem units <\tdim,\tdim>
\stpltsmbl
\photonrd -30 0 *5 /
\put {\large $W_3$} [r] at -35 0
\put {\large $B$} [l] at 35 0
\put {\EX} at 0 0
\linethickness=0pt
\put {\large Fig. 1d} at 0 -20
\putrule from 0 -15 to 0 15
\endpicture$$

$$\beginpicture
\setcoordinatesystem units <\tdim,\tdim>
\stpltsmbl
\photonrd -30 0 *2 /
\put {\large $Z$} [r] at -35 0
\put {\EX} at 0 0
\plot 30 15 0 0 30 -15 /
\tarrow from 20 10 to 16 8
\tarrow from 16 -8 to 20 -10
\put {\large $\overline{q}$} [l] at 35 15
\put {\large $q$} [l] at 35 -15
\linethickness=0pt
\putrule from 0 -45 to 0 45
\put {\large Fig. 2a} at 0 -50
\endpicture$$

$$\beginpicture
\setcoordinatesystem units <\tdim,\tdim>
\stpltsmbl
\circulararc 360 degrees from 20 0 center at 0 0
\photonrd -50 0 *2 /
\tarrow from 2 20 to -2 20
\tarrow from -2 -20 to 2 -20
\put {\large $t$} at 0 30
\put {\large $t$} at 0 -30
\put {\large $Z$} [r] at -55 0
\put {\EX} at 20 0
\plot 50 15 20 0 50 -15 /
\tarrow from 40 10 to 36 8
\tarrow from 36 -8 to 40 -10
\put {\large $\overline{q}$} [l] at 55 15
\put {\large $q$} [l] at 55 -15
\linethickness=0pt
\put {\large Fig. 2b} at 0 -50
\putrule from 0 -45 to 0 45
\endpicture$$

\newbox\phrdr
\setbox\phrdr=\hbox{\beginpicture
\setcoordinatesystem units <\tdim,\tdim>
\stpltsmbl
\setquadratic
\startrotation by .6 -.8 about 0 0
\plot
0 0
2.5 -3
5 0
7.5 3
10 0
/
\stoprotation
\endpicture}
\def\photonrdr #1 #2 *#3 /{\multiput {\copy\phrdr}  at
#1 #2 *#3 10 0 /}
$$\beginpicture
\setcoordinatesystem units <\tdim,\tdim>
\stpltsmbl
\plot -30 0 0 0 18 24 /
\tarrow from -17 0 to -13 0
\tarrow from 7.8 10.4 to 10.2 13.6
\put {\EX} at 0 0
\put {\large $b$} [r] at -35 0
\put {\large $s$} [l] at 23 24
\put {\large $\gamma$} [l] at 23 -24
\startrotation by .6 -.8 about 0 0
\photonrdr 0 0 *2 /
\stoprotation
\linethickness=0pt
\put {\large Fig. 3a} at 0 -50
\putrule from 0 -55 to 0 40
\endpicture$$
$$\beginpicture
\setcoordinatesystem units <\tdim,\tdim>
\stpltsmbl
\plot -60 0 0 0 36 48 /
\tarrow from -47 0 to -43 0
\tarrow from 7.8 10.4 to 10.2 13.6
\tarrow from 25.8 34.4 to 28.2 37.6
\put {\EX} at -15 0
\put {\large $b$} [r] at -65 0
\put {\large $s$} [l] at 41 48
\put {\large $\gamma$} [l] at 23 -24
\put {\large $t$} at 2 17
\put {\large $W$} at -25 31
\setquadratic
\plot
 18.0  24.0
 17.3  28.1
 13.3  26.9
  9.7  25.2
  8.1  28.9
  6.0  32.5
  2.7  29.9
 -0.1  27.0
 -2.8  29.9
 -6.1  32.4
 -8.3  28.8
 -9.8  25.2
-13.4  26.8
-17.4  28.0
-18.1  23.9
-18.2  19.9
-22.2  20.2
-26.4  19.9
-25.5  15.8
-24.2  12.0
-28.0  10.8
-31.7   9.0
-29.5   5.5
-26.9   2.5
-30.0  -0.0
/
\startrotation by .6 -.8 about 0 0
\photonrdr 0 0 *2 /
\stoprotation
\linethickness=0pt
\put {\large Fig. 3b} at 0 -50
\putrule from 0 -55 to 0 70
\endpicture$$

\end{document}